\documentclass[aps,prd,twocolumn,groupedaddress,showkeys,showpacs,superscriptaddress]{revtex4}
\usepackage{graphicx}
\usepackage{dcolumn}
\usepackage{bm}
\usepackage{natbib}
\usepackage{multirow}
\usepackage{epsfig}
\usepackage{amsmath}

\newcommand{\aap}{{Astron.~Astrophys.}}
\newcommand{\apjl}{{Astrophys.~J.~Lett.}}
\newcommand{\apjs}{{Astrophys.~J.~Supp.}}
\newcommand{\aj}{{Astron.~J.}}
\newcommand{\mnras}{{Mon.~Not.~R.~Astron.~Soc.}}
\newcommand{\physrep}{{Phys.~Rep.}}
\newcommand{\jcap}{{J. Cosm. Astropart. Phys.}}

\newcommand{\beq}{\begin{equation}}
\newcommand{\eeq}{\end{equation}}
\newcommand{\beqa}{\begin{eqnarray}}
\newcommand{\eeqa}{\end{eqnarray}}

\newcommand{\nhat}{\hat{\bf n}}

\newcommand{\be}{\begin{equation}}
\newcommand{\ee}{\end{equation}}

\newcommand{\rmd}{{\rm d}}

\newcommand{\kvec}{{\bf k}}
\newcommand{\khat}{\hat{\bf k}}

\newcommand{\fnl}{\tilde{f}_{\rm NL}}

\begin{document}

\title{Non-Gaussianity and large-scale structure in a two-field inflationary model}

\author{Dmitriy Tseliakhovich}
\author{Christopher Hirata}
\affiliation{Caltech M/C 350-17, Pasadena, California 91125, USA}
\author{An\v{z}e Slosar}
\affiliation{Brookhaven National Laboratory, Upton NY 11973, USA}

\date{\today}

\begin{abstract}
Single field inflationary models predict nearly Gaussian initial conditions and hence a detection of non-Gaussianity would be a signature of the more 
complex inflationary scenarios. In this paper we study the effect on the cosmic microwave background and on large scale structure from primordial non-Gaussianity in 
a two-field inflationary model in which both the inflaton and curvaton contribute to the density perturbations.  We show that in addition to the previously described 
enhancement of the galaxy bias on large scales, this setup results in large-scale stochasticity.  We provide joint constraints on the local non-Gaussianity parameter 
$\tilde f_{\rm NL}$ and the ratio $\xi$ of the amplitude of primordial perturbations due to the inflaton and curvaton using WMAP and SDSS data.
\end{abstract}

\pacs{98.80.Cq}
\keywords{Non-Gaussianity, Inflation} 

\maketitle

\setcounter{footnote}{0}

\section{Introduction}

One of the most important questions that cosmology faces today is the origin of structure in the universe. The generally accepted paradigm is that of inflation~\cite{Infl1,Infl2,Infl3,Infl4} which 
produces small adiabatic perturbations that evolve into the observed structure. The inflationary paradigm is extremely powerful as it remedies most of the problems of 
the original Big Bang scenario and also has a set of predictions that are well confirmed by current observations. On the other hand, although the generic predictions 
of inflation are quite clear, the nature of specific physical processes that govern inflation are still poorly understood.

The major obstacle in understanding inflation is that it can not be directly observed either in the laboratory or with telescopes. This problem is at the same time a 
virtue of inflation as it allows to indirectly probe physics at energies and time-scales that are far beyond the reach of current facilities. By comparing 
astrophysical observations with predictions of various inflationary models one can expect to distinguish between different extensions of the Standard Model of 
particle physics~\cite{LythReview}. Understanding of the reheating phase of inflation can provide a link between scalar fields driving inflation and the observable Universe that 
consists of dark and baryonic matter.

One of the many possible ways to deeper understand inflation is by studying the primordial density fluctuations. The usual inflationary model of a slowly-rolling 
inflaton field requires that the perturbations are highly Gaussian~\cite{Gaus1,Gaus2,Gaus3,Gaus4} and hence the detection of non-Gaussianity in either the cosmic 
microwave background (CMB) spectrum or the large scale structure (LSS) distribution would be a clear evidence that the physics driving inflation is more complicated 
than the standard inflaton scenario.

Non-Gaussianity naturally arises in inflationary models with more than one field~\cite{MField1,MField2,MField3,MField4}. One of the most studied models is the curvaton model~\cite{MField4,Curv1,Curv2,Curv3,Curv4,Curv5}, in which initial perturbations 
are generated by the curvaton field after inflation is over. In this model significant non-Gaussianity can be generated since the predicted 
curvature perturbation is proportional to the square of the curvaton field (as distinct from single-field inflation, where the required smoothness of the inflaton 
potential renders the curvature perturbation very nearly linear in the field fluctuations).

Most attempts to constrain non-Gaussianity have used the so-called ``local-type'' or $f_{\rm NL}$ parameterization~\cite{Komatsu00} in which one includes a quadratic 
term into the 
primordial potential, $\Phi = \phi + f_{\rm NL}\phi^2$.  In this parametrization both linear and quadratic terms in the potential originate from the same Gaussian 
field, e.g. a curvaton field, and the contributions from perturbations in other fields (e.g. the inflaton field responsible for inflation itself) are negligible.

The signature of local-type non-Gaussianity in the CMB has been described at length \cite{CMBNG1}.
It has also been established that $f_{\rm NL}$ has an effect on the galaxy bispectrum \cite{Verde00,Sefusatti07,Jeong09}.
The effect on the large-scale galaxy power spectrum has 
been considered only recently \cite{Dalal08, Slosar08, Carbone08, Afshordi08, McDonald08}, but it rapidly became clear that the method was competitive, stimulating 
work on $N$-body simulations of halo formation in non-Gaussian cosmologies \cite{Desjacques09, Grossi, P10, Reid}.
Recent constraints have been derived from the CMB bispectrum as measured by WMAP \cite{Komatsu03, Spergel07, Yadav08, Komatsu08, Komatsu10, Smith09,i3} 
and from large scale structure in the 
Sloan Digital Sky Survey (SDSS) \cite{Slosar08}.  Recently, $\sim3\sigma$ evidence for excess clustering consistent with non-Gaussianity has been 
identified in the NRAO VLA Sky Survey (NVSS) \cite{XiaNVSS}.

In this paper we extend the formalism to include the case where both the inflaton and curvaton contribute significantly to the curvature perturbation.  Perturbations generated by the inflaton field are 
purely Gaussian, while curvaton fluctuations can result in non-Gaussianity if the conversion from curvaton fluctuation $\delta\sigma$ to primordial potential $\Phi$ contains quadratic terms.  The ratio 
of inflaton to curvaton contributions $\xi$ is arbitrary: the framework of the curvaton model allows it to take on any positive value.  Usually one takes $\xi\gg 1$ since in the opposite limit ($\xi\ll 
1$) the curvaton has no observable effect on the primordial perturbations.  Here we investigate the consequences of general $\xi$ -- including values of order unity -- for the CMB and LSS.  The type of 
non-Gaussianity generated could be called ``local-stochastic,'' in that it results from local nonlinear evolution of the inflaton and curvaton fields (and thus the primordial bispectrum will have the 
local-type configuration dependence), but that the full nonlinear potential $\Phi$ is not a deterministic function of the linear potential.

Studying non-Gaussianity is particularly important in the face of the current generation of CMB projects~\cite{CMBTF} such as the {\slshape Planck} satellite as well 
as ongoing and future LSS projects.  To fully exploit the potential of the future probes it is imperative to investigate theoretically the range of types of 
non-Gaussianities that can be produced in unconventional inflation (e.g. multi-field models), and understand what effect they have on the CMB and LSS.

The rest of the paper is organized as follows. In Sec.~\ref{sec:theory} we discuss the generation of non-Gaussian primordial perturbations in the inflationary model 
with both inflaton and curvaton fields contributing to the curvature perturbation.
In Sec.~\ref{sec:cmb} we describe the effect of two-field models on the CMB bispectrum.
In Sec.~\ref{sec:halo} we derive the halo power spectrum using the peak-background
split formalism \cite{Cole}, and in Sec.~\ref{sec:lss} we consider the angular 
power spectrum of galaxies. Section~\ref{sec:constraints} provides the constraints on the two-field model from existing data, and we conclude in Sec.~\ref{sec:disc}.


\section{Non-Gaussian initial perturbations in two-field inflationary models}
\label{sec:theory}

We consider a model of inflation where both the inflaton {\em and} the curvaton contribute to the primordial density perturbations.  This configuration can exhibit a rich set of phenomenology, including 
both non-Gaussianity and various mixtures of adiabatic and isocurvature perturbations \cite{a0403258, a0407300, h0409335, LVW, 1004.0818}.  In this paper, we will restrict ourselves to the 
case where the dark matter decouples after the curvaton decays and its energy density is thermalized.  This ensures that no dark matter isocurvature perturbation is produced, and the only observable 
perturbation is the curvature perturbation $\zeta$ that is conserved between curvaton decay and horizon entry.

The simplest case is that of two non-interacting scalar fields: the inflaton $\varphi$ and the curvaton $\sigma$.  The latter is taken to have a quadratic potential,
\beq
V_\sigma(\sigma) = 
\frac12m^2\sigma^2.
\label{eq:Vsigma}
\eeq
During inflation, the inflaton dominates the energy density of the Universe, whereas the curvaton is effectively massless ($m\ll H$) and pinned by Hubble friction to a fixed value $\bar\sigma$ (aside 
from perturbations to be described later).  Quantum fluctuations generate a spectrum of perturbations
$\delta\varphi$ and $\delta\sigma$ in both the inflaton and curvaton fields:
\beq
P_{\delta\varphi}(k) = \frac{H_*^2}{2k^3} \qquad {\rm and} \qquad P_{\delta\sigma}(k) = \frac{H_*^2}{2k^3},
\eeq
where $H_*$ is the Hubble rate evaluated at the horizon crossing for a given mode, i.e. when $k = aH$, and the post-horizon-exit field perturbations are defined on the uniform total density slice.  
The $\varphi$ and $\sigma$ perturbations are nearly Gaussian and uncorrelated.

The inflaton perturbation is parallel to the unperturbed trajectory in $(\varphi,\sigma)$-space and hence is an adiabatic perturbation; indeed it behaves the same way that perturbations behave in 
single-field inflation.
The curvaton perturbation 
however is an isocurvature perturbation and can have complicated dynamics.  In the 
simplest version of the curvaton scenario, the curvaton begins to 
oscillate after the end of inflation when the Hubble rate drops to $H\sim m$.  As a massive scalar with zero spatial momentum, its energy density subsequently redshifts as $\rho_\sigma\propto a^{-3}$.  
Of interest to us is the fact that for quadratic potentials (Eq.~\ref{eq:Vsigma}) this energy density is also proportional to the square of the curvaton field $\sigma=\bar\sigma + \delta\sigma$, i.e.
\beq
\delta_\sigma \equiv
\frac{\delta\rho_{\sigma}}{\bar\rho_{\sigma}}
= 2\frac{\delta\sigma}{\bar\sigma} + \frac{\delta\sigma^2-\langle\delta\sigma^2\rangle}{\bar\sigma^2},
\eeq
where the subtraction of the variance arises from the ${\cal O}(\delta\sigma^2)$ expansion of $\bar\rho_{\sigma}\propto\bar\sigma^2+\langle\delta\sigma^2\rangle$.
The quadratic term allows the curvaton to generate a non-Gaussian density perturbation.  In the radiation-dominated era, the curvaton's contribution to the energy density increases as 
$\Omega_\sigma\propto a$, thereby enhancing the importance of $\delta_\sigma$.  The decay of the curvaton and the thermalization of its energy density result in a non-Gaussian adiabatic perturbation.

The $\delta N$ formalism \cite{dN1} extended into the nonlinear regime \cite{Gaus4} quantitatively provides the curvature perturbation to second order in the field 
perturbations 
$(\delta\varphi,\delta\sigma)$; this is (Eq.~26 of Langlois {\slshape et~al.}~\cite{LVW})
\beq
\zeta = -\frac{H_* \delta\varphi}{\dot\varphi_*} + \frac{2r}{3}\frac{\delta\sigma}{\sigma_*}
  + \frac{2r}{9}\left(\frac32-2r-r^2\right)\frac{\delta\sigma^2}{\sigma_*^2},
\eeq
where the subscript $_*$ denotes evaluation at horizon exit, and $r$ is related to the fraction of the energy density in the curvaton when it decays:
\beq
r = \left. \frac{3\rho_\sigma}{4\rho_{\rm rad} + 3\rho_\sigma} \right|_{\rm decay}.
\label{eq:rdef}
\eeq

The primordial potential perturbation in the Newtonian gauge is then given by the usual expression $\Phi=\frac35\zeta$
(e.g. \cite{LL}; but note that in large scale structure non-Gaussianity studies the opposite sign 
convention is adopted, so that $\Phi>0$ corresponds to overdensities).

We may put the primordial potential in a form more closely related to that of large-scale structure non-Gaussianity studies:
\beq
\Phi({\bf x}) = \phi_1({\bf x}) + \phi_2({\bf x}) + \fnl[ \phi_2^2({\bf x})-\langle\phi_2^2\rangle ],
\label{eq:Phi}
\eeq
where $\phi_1$ and $\phi_2$ are the parts of the linear primordial potential corresponding to the inflaton and curvaton fields respectively.
Their power spectra are given by
\beq
\frac{k^3}{2\pi^2} P_{\phi_1}(k) = \frac9{25} \left( \frac {H_*^2}{2\pi\dot\varphi_*} \right)^2
\label{phi1}
\eeq
and
\beq
\frac{k^3}{2\pi^2} P_{\phi_2}(k) = \frac{4r^2}{25} \left( \frac{H_*}{2\pi\sigma_*} \right)^2.
\eeq
The non-Gaussianity parameter is
\beq
\fnl = \frac{5}{6r}\left(\frac32-2r-r^2\right).
\label{eq:fnl}
\eeq
(We use the tilde since the label ``$f_{\rm NL}$'' is usually used to denote the non-Gaussianity parameter appearing in the primordial bispectrum.)

It is convenient to specify the relative contribution of the inflaton and curvaton fields to the primordial potential using the
ratio of standard deviations $\xi = \sigma(\phi_1)/\sigma(\phi_2)$.  Thus a fraction $\xi^2/(1+\xi^2)$ of the power comes from the
inflaton and a fraction $1/(1+\xi^2)$ from the curvaton.  This ratio is
\beq 
\xi(k) = \left| \frac{3\sigma_*H_*}{2r\dot\phi_*} \right| = \frac3{2r} \left| \frac{\sigma_*}{(d\phi/dN)_*} \right|,
\label{eq:xi}
\eeq
where $N$ is the number of $e$-folds remaining in inflation.
Thus the observable features of this model are specified by the primordial power spectrum $P_\Phi(k)$ and by the two new parameters $\fnl$ and $\xi$ (in principle $\xi$ will have a scale dependence
$d\ln\xi/d\ln k$  of order the slow roll parameters, but unless non-Gaussianity is detected at high statistical significance this cannot be measured).  We will work with these parameters from here 
forward.

\section{The CMB bispectrum}
\label{sec:cmb}

The effect of local-type non-Gaussianity on the CMB bispectrum has a long history, both in purely adiabatic models as considered here, and in locally non-Gaussian isocurvature models \cite{i1,i2}.  
We evaluate the CMB bispectrum using our set of parameters here.

The CMB constraints on primordial non-Gaussianity come from the measurements of the CMB angular bispectrum~\cite{Komatsu00},
\begin{equation}
  \label{eq:blllmmm}
  B_{\ell_1\ell_2\ell_3}^{m_1m_2m_3}\equiv 
  \left\langle a_{\ell_1m_1}a_{\ell_2m_2}a_{\ell_3m_3}\right\rangle,
\end{equation}
where $a_{\ell m}$ is the CMB temperature fluctuation expanded in spherical harmonics:
\begin{equation}
  a_{\ell m}\equiv \int d^2\hat{\mathbf n}\frac{\Delta T(\hat{\mathbf n})}{\bar T}
  Y_{\ell m}^*(\hat{\mathbf n}).
\end{equation}
If the primordial fluctuations are adiabatic scalar fluctuations, then $a_{\ell m}$ can be easily expressed in terms of the primordial potential $\Phi$ and the radiation transfer function $g_\ell(k)$:
\begin{equation}
  a_{\ell m}=4\pi(-i)^\ell
  \int\frac{d^3{\mathbf k}}{(2\pi)^3}\Phi({\mathbf k})g_\ell(k)
  Y_{\ell m}^*(\hat{\mathbf k}).
\end{equation}
From Eq.~(\ref{eq:Phi}) it follows that in the Fourier space primordial potential can be decomposed into
parts associated with the linear potential perturbations $\phi_1$ and $\phi_2$, and with the nonlinear coupling $\fnl$:
\beq
\Phi({\mathbf k})=\phi_1({\mathbf k}) + \phi_2({\mathbf k}) + \phi_{NL}({\mathbf k}).
\eeq
Here $\phi_{NL}({\mathbf k})$ is the $\fnl$-dependent part,
\begin{equation}
  \label{eq:nonlinear}
  \phi_{NL}({\mathbf k})\equiv 
  \fnl
  \int \frac{d^3{\mathbf p}}{(2\pi)^3}
  \phi_2({\mathbf k}+{\mathbf p})\phi^\ast_2({\mathbf p}).
\end{equation}
Using Wick's theorem we calculate the bispectrum of the total potential, $B_\Phi(k_1,k_2,k_3)$.  It contains one contribution from allowing each of
the $\Phi({\mathbf k}_i)$ to have a contribution from $\fnl$; in the case where this is $k_3$:
\beqa
  \nonumber
  \left\langle[\phi_1({\mathbf k}_1) + \phi_2({\mathbf k}_1)]
	[\phi_1({\mathbf k}_2) + \phi_2({\mathbf k}_2)]
	\phi_{NL}({\mathbf k}_3)\right\rangle = \\ 
	2(2\pi)^3\delta^{(3)}({\mathbf k}_1+{\mathbf k}_2+{\mathbf k}_3)
    \frac{\fnl}{(1 + \xi^2)^2}P_\Phi(k_1)P_\Phi(k_2),
\eeqa
where $P_\Phi(k)$ is the power spectrum of the total potential.  The total bispectrum is the sum of this and the similar contributions where $\fnl$
contributes to ${\mathbf k}_1$ and to ${\mathbf k}_2$:
\beqa
B_\Phi(k_1,k_2,k_3) &=& 2\frac{\fnl}{(1 + \xi^2)^2} [
P_\Phi(k_1)P_\Phi(k_2) + 
\nonumber \\ &&
\hspace{-1.5cm} P_\Phi(k_1)P_\Phi(k_3)+ P_\Phi(k_2)P_\Phi(k_3)]. 
\label{eq:bphi}
\eeqa

For constraining non-Gaussianity it is convenient to introduce two new variables: $x_1 = \fnl/(1+\xi^2)^2$ and $x_2 = 1/(1 + \xi^2)$.  Using the
bispectrum of $\Phi({\mathbf k})$,
we can finally write the CMB angular bispectrum (via a calculation similar to Ref.~\cite{Komatsu00}):
\begin{eqnarray}
  \nonumber
  B_{\ell_1\ell_2\ell_3}^{m_1m_2m_3} &=& 
  2{\cal G}_{\ell_1\ell_2\ell_3}^{m_1m_2m_3}
	\int_0^\infty r^2 dr 
\nonumber \\ && \times
    [b^L_{\ell_1}(r)b^L_{\ell_2}(r)b^{NL}_{\ell_3}(r)
 + b^L_{\ell_1}(r)b^{NL}_{\ell_2}(r)b^{L}_{\ell_3}(r)
\nonumber \\ &&
 + b^{NL}_{\ell_1}(r)b^L_{\ell_2}(r)b^{L}_{\ell_3}(r)],
\end{eqnarray}
where ${\cal G}_{\ell_1\ell_2\ell_3}^{m_1m_2m_3}$ is the Gaunt integral, and we use
\begin{equation}
  \label{eq:bLr}
  b^L_{\ell}(r) \equiv
  \frac2{\pi}\int_0^\infty k^2 dk\, P_\Phi(k)g(k)j_\ell(kr)
\end{equation}
and
\begin{equation}
  \label{eq:bNLr}
  b^{NL}_{\ell}(r) \equiv
  \frac{2x_1}{\pi}\int_0^\infty k^2 dk\, g(k)j_\ell(kr).
\end{equation}

We see from Eq.~(\ref{eq:bphi}) that CMB bispectrum measurements of $f_{\rm NL}^{\rm loc}$ that assume a purely curvaton contribution ($\xi=0$) are actually measuring 
$x_1$ in the more general case.

\begin{figure}
\vspace{20pt}
\includegraphics[width=3.4in]{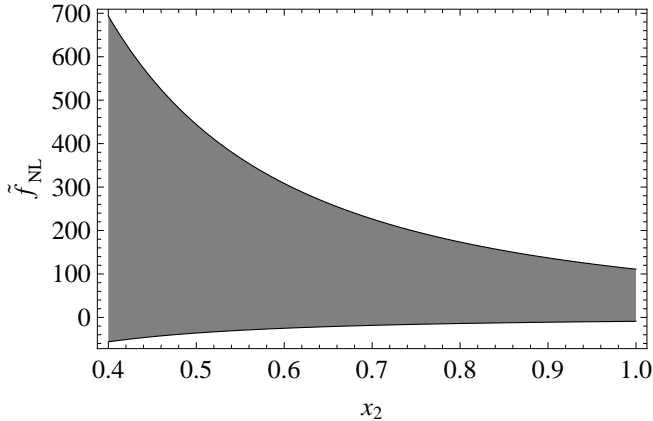}
\caption{\label{fnl1}The allowed range ($2\sigma$) of $\fnl$ as a function of $x_2$ derived from the WMAP data~\cite{Komatsu08}. As discussed in the text, $\tilde f_{NL}$ becomes unconstrained as $x_2 \rightarrow 0$ because in this case the statistics describing the density distribution is dominated by the inflaton field and is nearly Gaussian.}
\end{figure}

Clearly, when the curvaton field dominates the perturbation power ($x_2 \approx 1$), we have $f_{\rm NL}=\fnl$, and
our $\fnl$ constraints are identical to the constraints on models with negligible inflaton perturbation.  However, as the 
contribution from the inflaton field increases ($x_2 \rightarrow 0$) the role of the contribution of the curvaton field to the primordial curvature perturbation becomes negligible and the statistics describing the density distribution becomes very nearly Gaussian. In that case $\fnl$ becomes completely unconstrained (Fig.~\ref{fnl1}) and we can no longer make robust predictions regarding the presence of the second field on the basis of the non-Gaussianity observations alone. The CMB bispectrum is therefore not capable of breaking the degeneracy between $\fnl$ and $\xi$.  The use of other constraints is necessary.  In this paper, we will use large-scale structure, although we note that in principle the CMB trispectrum might also be useful for this purpose, since it scales as $\fnl^2/(1+\xi^2)^3 = x_1^2/x_2$ and thus in combination with the bispectrum would allow the $(\fnl,\xi)$ degeneracy to be broken.

\section{Local non-Gaussianity in peak-background split formalism}
\label{sec:halo}

In this section we outline the inflationary scenario in which both inflaton and curvaton fields are contributing to the initial curvature perturbation and derive the 
power spectrum of galaxies using the peak-background split formalism~\cite{Cole}.

We decompose the density field into a long-wavelength and short-wavelength pieces:
\beq
\rho(\mathbf{x}) = \bar{\rho}(1 + \delta_l + \delta_s).
\eeq
The linear density field here is a sum of two independent Gaussian components $\delta = \delta^{(1)} + \delta^{(2)}$ originating from the inflaton and curvaton fields, respectively.

The local Lagrangian density of halos now depends on the large scale matter perturbations of both gaussian fields:
\beq
n(\mathbf{x}) = \bar{n}[1 + b_1\delta_l^{(1)} + b_2\delta_l^{(2)}].
\label{eq:nb1}
\eeq
We will see that $b_i$ can be scale-dependent ($k$-dependent), which means that in position space it should be thought of as a (possibly nonlocal) 
operator acting on the density field, 
i.e. $b_2\delta_l^{(2)}(\mathbf{x})$ is the convolution of $\delta_l^{(2)}(\mathbf{x})$ with the Fourier transform of $b_2(\mathbf{k})$.

We can express the bias parameters in terms of the number density function
\beq
b_i = \bar{n}^{-1}\frac{\partial n}{\partial \delta_l^{(i)}}.
\eeq
It is easy to check that the bias for the $\delta_1$ field is just the usual Lagrangian bias that applies to Gaussian cosmologies;
for example, in the Press-Schechter model~\cite{PS} it is $b_1 = 
b_g \equiv \delta_c/\sigma_{\delta}^2-\delta_c^{-1}$ with $\delta_c=1.686$ quantifying the spherical collapse linear over-density. To calculate $b_2$ 
we note that 
short-wavelength 
modes $\delta_s$ in an overdense region determined by $\delta_l$ can be written as:
\beqa
\delta_s = \alpha \left[ (1+2\fnl\phi_l^{(2)})\phi_s^{(2)} + \fnl(\phi_s^{(2)})^2 + \phi_s^{(1)} \right]
\nonumber\\
\equiv \alpha \left[ X_1\phi_s^{(2)} + X_2(\phi_s^{(2)})^2 + \phi_s^{(1)} \right], \label{eq:ds}
\eeqa
where $X_1=1+2\fnl\phi_l^{(2)}$ and $X_2=\fnl$.  Here $\alpha$ is the transfer function that converts the potential into the density 
field, $\delta(k) = \alpha(k)\Phi(k)$.  In general one may think of $\alpha$ as an operator defined by its action in Fourier space, i.e. when
applied to a real-space function such as $\phi({\bf x})$, we have
\beq
\alpha\phi({\bf x}) \equiv \int \frac{d^3{\bf k}}{(2\pi)^3} \alpha(k) e^{i{\bf k}\cdot{\bf x}} \int d^3y\,e^{-i{\bf k}\cdot{\bf y}} \phi({\bf y}).
\eeq
The specific function $\alpha(k)$ is given by Eq.~(7) of Slosar {\it et~al.} \cite{Slosar08}:
\beq
\alpha(k;z) = \frac{2c^2k^2T(k)D(z)}{3\Omega_mH_0^2},
\eeq
where $T(k)$ is the linear transfer function with conventional normalization $T(0)=1$, and $D(a)$ is the growth function normalized to $D(z)=
(1+z)^{-1}$ at high redshift.  The inverse operator $\alpha^{-1}$ is obtained by the replacement $\alpha(k)\rightarrow\alpha^{-1}(k)=1/\alpha(k)$.

This shows that local number density in the non-Gaussian case depends not only on $\delta_l$, but also on $X_1$, $X_2$, and hence $b_2$ becomes
\begin{equation}
b_2 = \bar n^{-1} \left[ \frac{\partial n}{\partial \delta_l^{(2)}({\bf x})}
  + 2\fnl \alpha^{-1} \frac{\partial n}{\partial X_1}
  \right],\label{eq:hb1}
\end{equation}
where the derivative is taken at the mean value $X_1=1$.

We can further simplify this expression by considering a rescaling of the power spectrum on 
the small scales 
due to the presence of non-Gaussianity.  In a ``local'' region of some size $R$, and for small-scale Fourier modes $k\gg R^{-1}$ within this region, 
there is a local power spectrum
\beqa
P_s^{\rm local}(k) &=& \frac{\xi^2 + (1 + 2\fnl\alpha^{-1}\delta_l^{(2)})^2}{1+\xi^2} P_s^{\rm global}(k)
\nonumber \\ &\equiv& X_0P_s^{\rm global}(k),
\eeqa
from which we obtain the rescaling of $\sigma_8$:
\beq
\sigma_8^{\rm local} = \sigma_8 \sqrt{X_0}.
\eeq
Using these expressions we can change the derivatives in Equation (\ref{eq:hb1}) to finally obtain
\begin{equation}
b_2(k) = b_{\rm g} + \frac{2\fnl}{1+\xi^2} \alpha^{-1}(k) \frac{\partial \ln{n}}{\partial \ln{\sigma_8^{\rm local}}}.
\label{eq:hb2}
\end{equation}

For further calculations we assume the mass function to be universal, i.e. we assume that it depends only on the significance function $\nu(M)\equiv 
\delta_{\rm c}^2/\sigma^2(M)$:
\begin{equation}
n(M,\nu)= M^{-2} \nu f(\nu) \frac{\rmd\ln\nu}{\rmd\ln M}.
\end{equation}
This assumption is much more general than the Press-Schechter picture, e.g. it holds for the Sheth-Tormen mass function \cite{S-T} as well.  Universality implies that
\beq
\frac{\partial \ln{n}}{\partial \ln{\sigma_8^{\rm local}}} = \delta_{\rm c}b_{\rm g},
\eeq
from which we derive
\beq
b_2(k) = b_{\rm g} + \frac{2\delta_{\rm c}\fnl}{1+\xi^2}b_{\rm g} \alpha^{-1}(k).
\label{eq:b2}
\eeq

The standard Gaussian bias in Eulerian space is given by $b\equiv b_{\rm g}+1$.  The halo overdensity in Eulierian space in the non-Gaussian case is then obtained by 
multiplying Eq.~(\ref{eq:nb1}) by $1+\delta_l$; to first order,
\beqa \label{deltah}
\delta_{\rm h} &\equiv& \frac{n({\mathbf x})}{\bar n}-1 = \delta_l + b_1\delta_l^{(1)} + b_2\delta_l^{(2)}
\nonumber \\
&=& [1+b_1(k)]\delta_l^{(1)} + [1+b_2(k)]\delta_l^{(2)}.
\eeqa
We can now write down the halo power spectrum in the form:
\beq
P_{\rm hh}(k) = \frac{(1+b_1)^2\xi^2 + [1+b_2(k)]^2}{1+\xi^2} P^{\rm lin}(k).
\eeq
Finally, plugging in $b_1(k)=b$ and using Eq.~(\ref{eq:b2}) for $b_2(k)$, we obtain
\beqa \label{Pgeq}
P_{\rm hh}(k) &=&
\frac{\xi^2b^2 + [b + 2(b-1)\fnl\delta_c(1+\xi^2)^{-1}\alpha^{-1}(k)]^2 }{1+\xi^2}
\nonumber \\ && \times P^{\rm lin}(k).
\eeqa
In the limit of $\xi \rightarrow 0$, i.e. when the contribution from the inflaton field is negligible
we recover the standard expression for the power spectrum with the curvaton generated non-Gaussianity (Eq.~32 of~\cite{Slosar08}).

(It should be noted that the non-Gaussianity also introduces small corrections to the scale-independent part of the bias, because the small-scale 
fluctuations that must collapse to form a massive halo have a non-Gaussian density distribution.  This effect has been seen in simulations with 
$f_{\rm NL}$-type non-Gaussianity, where it is negative for $f_{\rm NL}>0$, resulting in a slight reduction of the non-Gaussian bias enhancement at 
large $k$, and even a change in sign of the $f_{\rm NL}$ effect at very small scales \cite{Desjacques09,P10}.  However, since current studies of 
non-Gaussianity using LSS allow the scale-independent bias to be a free parameter, they are not sensitive to this effect; it would only be important 
if the Gaussian contribution to the bias were inferred indepedently, e.g. from measurements of halo mass and the mass-bias relation.)

\begin{figure}
\includegraphics[angle=-90,width=3.2in]{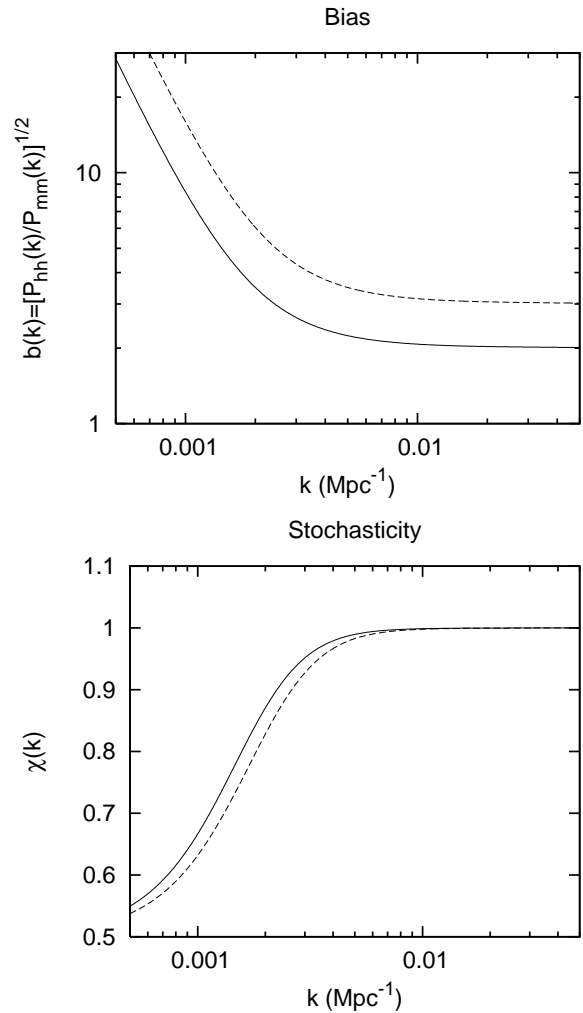}
\caption{\label{fig:stoch}The bias and stochasticity of galaxies at $z=1$ in a model with $x_1=30$ and $x_2=0.5$ ($\fnl=120$).  
The solid lines show a tracer with $b=2$ and the dashed lines a tracer with $b=3$.
The background cosmology and power spectrum are those of WMAP5.}
\end{figure}

A further consequence of this model that does not arise in the case with only curvaton perturbations is large-scale stochasticity.  In particular, the squared correlation coefficient
\beq
\chi(k) = \frac{P_{\rm hm}^2(k)}{P_{\rm hh}(k)P_{\rm mm}(k)}
\eeq
deviates from unity on the largest scales.  We can see this by writing
the cross power spectrum $P_{\rm hm}(k)$ as
\beq
P_{\rm hm}(k) = \frac{(1+b_1)\xi^2 + [1+b_2(k)]}{1+\xi^2} P^{\rm lin}(k).
\eeq
In the linear Gaussian theory, one would have $\chi=1$, whereas in our case we have
\beq
\chi(k) = \frac{\{(1+b_1)\xi^2 + [1+b_2(k)]\}^2}{(1+\xi^2)\{(1+b_1)^2\xi^2 + [1+b_2(k)]^2\}}.
\label{eq:chi-k}
\eeq
Note that if $\fnl\neq0$, on large scales $|b_2(k)|\gg 1,b_1$ and hence
\beq
\lim_{k\rightarrow 0^+} \chi(k) = \frac1{1+\xi^2} = x_2.
\eeq
An example of the onset of scale-dependent bias and stochasticity is shown in Figure~\ref{fig:stoch}; note that this type of stochasticity effect exists only for $x_2\neq1$.

It is important to note that stochasticity can arise even for $x_2=1$ in two ways.  One is that on small scales, there is a breakdown of linear biasing. However, since our constraints on non-Gaussianity arise from the largest scales in the survey (mainly the $l<25$ quasar data points) this effect can be neglected.  The other is halo shot noise (e.g. \cite{Seljak}), which arises from the fact that haloes containing multiple galaxies (or quasars) can produce a large ``1-halo'' contribution to the correlation function at small separations.  When transformed to Fourier space at large scales (small $k$), this results in additional contribution to the power spectrum of
\begin{equation}
\lim_{k\rightarrow 0^+} P_{\rm 1~halo}(k) =
\int 4\pi r^2\xi_{\rm 1~halo}(r)\,dr,
\label{eq:corrfuncint}
\end{equation}
where $\xi_{\rm 1~halo}(r)$ is one-halo correlation function. In principle since $P(k)\propto k$ on large scales, the halo shot noise term ($\propto k^0$) may become important; since it is random and not determined by the underlying long-wavelength modes of the density field, it also produces stochasticity.

However, the halo shot noise is expected to be a very small contribution for our data sets.  A simple way to see this is to note that the ratio of the halo shot noise to the usual shot noise is equal to twice the ratio of 1-halo pairs to the number of galaxies (this follows from Eq.~(\ref{eq:corrfuncint}) and the definition of the correlation function).  For the quasar sample, a simple counts-in-pixels analysis of the catalog suggests that 0.6\% of the quasars are in pairs (the Healpix pixels \cite{Gorski} used are 3.5 arcmin in size, i.e. much larger than haloes at $z>1$), suggesting that the halo shot noise term is $C_{l,{\rm 1 halo}}\sim 0.012/\bar n_{\rm 2D}$.  This is two orders of magnitude smaller than the error bars on the lowest-$l$ quasar autopower point displayed in Slosar {\slshape et~al.} \cite{Slosar08} and hence negligible.

\section{Galaxy angular power spectrum}
\label{sec:lss}

Additional constraints on the primordial non-Gaussianity come from the observations of the Large Scale Structure (LSS), and in particular from large galaxy surveys such as SDSS. These constraints are primarily driven by extremely large scales, corresponding to wave numbers $k < 0.01 Mpc^{-1}$. The data sets used for LSS constraints include spectroscopic and photometric Luminous Red Galaxies (LRGs) from SDSS, photometric quasars from SDSS and cross-correlation between galaxies and dark matter via Integrated Sach-Wolfe effect. Detailed description of the data used for LSS constraints can be found in~\cite{Ho,Slosar08} and here we will only emphasize the redshift ranges of the most important datasets we used.

The photometric LRGs dataset was constructed and discussed in detail in~\cite{Paddy05}. The  sample was sliced into two redshift bins with $0.2 < z < 0.4$ and $0.4 < z < 0.6$. Our spectroscopic LRG power spectrum comes from~\cite{Tegmark93} and is based on a galaxy sample that covers $4000$ square degrees of sky over the redshift range $0.16 < z < 0.47$. Quasars for our constraints were photometrically selected from the SDSS quasar catalog consisting of UVX objects~\cite{Richards} and DR3 catalog~\cite{Richards2}. Quasars in our sample fall into two redshift bins with $0.65 < z < 1.45$ and $1.45 < z < 2.0$.

The tightest LSS constraints on primordial non-Gaussianity involve purely photometric samples (where one observes the two-dimensional projection of the 
galaxy distribution with a set of color cuts) since this allows the largest volume to be probed with the highest number density.  At low $k$, where the effects of primordial non-Gaussianity on the power spectrum are largest, there is less of an 
advantage to having the large number of modes ($\propto k_{\rm max}^3$ instead of $k_{\rm max}^2$) achievable via spectroscopy.
In order to obtain constraints on the $\fnl$ and $\xi$ parameters from LSS we need to modify Eq.~(\ref{Pgeq}) to give the
angular power spectrum.

To obtain the angular power spectrum, we project the galaxy density field $\delta_{g,3D}$ along the line of sight ${\nhat}$ and take into account redshift distortions~\cite{AngPk1, AngPk2, Padmanabhan91}:
\beq
1+\delta_{g}(\nhat) = \int \, dy\, f(s) [1+\delta_{g,3D}(y, y\nhat)] \,\,.
\label{eq:deltag_red}
\eeq
Here, $s = y + H^{-1}{\mathbf v}\cdot\nhat$ is the redshift distance, $f(s)$ is the normalized radial selection function, and we have explicitly written the mean contribution to the density field. We 
further note that peculiar velocities are generally small compared to the size of the redshift slice and hence we can Taylor expand selection function as:
\beq
f(s) \approx f(y) + \frac1{aH} \frac{df}{dy} \,{\mathbf v}(y\nhat)\cdot\nhat \,\,.
\label{eq:fs}
\eeq
At large scales where the density perturbations are $\ll1$, we may ignore second-order terms, i.e. we may ignore the product of the velocity term in Eq.~(\ref{eq:fs}) with the $\delta_{g,3D}(y, y\nhat)$ 
term in Eq.~(\ref{eq:deltag_red}).
This allows us to split $\delta_g$ into two terms as $\delta_{g} = \delta_{g}^{0} + \delta_{g}^{r}$.  In terms of the Fourier transformed fields, we can write $\delta_g^0$ and $\delta_g^r$ as:
\beq
 \delta_{g}^0(\nhat) = \int \, dy \, f(y) \int\frac{d^{3}\kvec}{(2\pi)^3} \delta_{g,3D}(y, \kvec) e^{-i \kvec \cdot \nhat y} \,\,
\label{eq:deltagl1}
\eeq
and
\beq
\delta_{g}^{r}(\nhat) = \int \, dy\, \frac{df}{dy} \int \frac{d^{3} \kvec}{(2 \pi)^{3}} 
\frac1{aH}{\mathbf v}(y,\kvec) \cdot \nhat e^{-i \kvec \cdot \nhat y} \,\,.
\label{eq:deltag_v}
\eeq
The velocity can be related to the density perturbation using linearized continuity equation:
\beq
H^{-1}{\mathbf v}(y,\kvec) = -i \frac{\Omega_{m}^{0.6}}{b} \delta_{g}(y,\kvec) \frac{\kvec}{k^{2}}\,\,.
\label{eq:velocity}
\eeq

We can further transfer redshift dependence of $\delta_g$ into a growth function $D(y)$ and expand $\delta$'s in Legendre polynomials $P_{\ell}(x)$ using the following identity:
\beq
e^{-i \kvec \cdot \nhat y} = \sum_{\ell=0}^{\infty} (2\ell+1) i^{\ell} j_{\ell}(ky) P_{\ell}(\khat \cdot \nhat) \,\,,
\eeq
where $j_\ell(s)$ is the spherical Bessel function of order $\ell$. We obtain
\beq
\delta_{g}^0(\nhat) = \int\frac{d^{3}\kvec}{(2\pi)^3} \sum_{\ell=0}^{\infty} (2\ell+1) P_{\ell}(\khat \cdot \nhat)\delta_{g,\ell}^0 \,\,,
\eeq
where $\delta_{g,\ell}^0$ is the observed galaxy transfer function [analogous to the CMB radiation transfer function $g_\ell(k)$]:
\beq
\delta_{g,\ell}^0 = i^{\ell} \int dy \, f(y) \delta_{g,3D}(\kvec)D(y) 
j_{l}(ky). \,\,
\eeq
Similarly, we can write $\delta_{g,\ell}^r$ as
\beq
\delta_{g,\ell}^r = i^{\ell} \int dy \, \frac{df}{dy} \delta_{g,3D}(\kvec)D(y) 
\frac{\Omega_{m}^{0.6}}{kb} j'_{l}(ky) \,\,.
\eeq
Now we use Eq.~(\ref{deltah}) to express $\delta_{g,\ell}$ in terms of the overdensities generated by inflaton and curvaton fields: 
\beqa
\delta_{g,\ell}^0 = i^{\ell} \int dy \, f(y)D(y)( [1+b_1(k)]\delta_\ell^{(1)}
\nonumber \\
+ [1+b_2(k)]\delta_\ell^{(2)} ) j_{\ell}(ky)\hspace{1cm}
\eeqa
and
\beq
\delta_{g,\ell}^r = i^{\ell} \int dy \, \frac{df}{dy}D(y)\left(\delta_l^{(1)} + \delta_l^{(2)} \right)
\frac{\Omega_{m}^{0.6}}{k} j'_{l}(ky) \,\,.
\eeq

Using these expressions together with equation~(\ref{Pgeq}) it is straightforward to calculate angular power spectrum $C_\ell$, which can be conveniently divided into 
three terms:
\beq
C_\ell =  C^{gg}_\ell + C^{gv}_\ell +C^{vv}_\ell,
\eeq
corresponding to the pure real-space galaxy density contribution ($gg$), the redshift space distortion term ($vv$), and the cross-correlation ($gv$).
These components of the angular power spectrum are expressed in terms of the 3D linear matter power spectrum 
$\Delta_k^2$ and $x_1$, $x_2$ parameters, introduced in the previous section:
\begin{eqnarray}
C^{gg}_\ell &=& 4\pi\int \frac{\rmd k}k \,\Delta_k^2 (|W_\ell^0(k)|^2(1-x_2) + |W_\ell^1(k)|^2x_2),
\nonumber \\
C^{gv}_\ell &=& 8\pi\int \frac{\rmd k}k \,\Delta_k^2 \Bigl( \Re \left[ W_\ell^{0\ast}(k) W_\ell^r(k)\right](1 -x_2) + 
\nonumber \\
&+& \Re \left[ W_\ell^{1\ast}(k) W_\ell^r(k) \right]x_2 \Bigr), {\rm ~~and}
\nonumber \\
C^{vv}_\ell &=& 4\pi\int \frac{\rmd k}k \,\Delta_k^2 |W_\ell^r(k)|^2,
\label{eq:cgv}
\end{eqnarray}
where the window functions are given by
\begin{eqnarray}
W^0_\ell(k) &=& \int bD(y)f(y) j_\ell(ky)\,\rmd y,
\nonumber \\
W^1_\ell(k) &=& \int (b+\Delta b)D(y)f(y) j_\ell(ky)\,\rmd y, {\rm ~~and}
\nonumber \\
W_\ell^r(k) &=& \int \Omega_m^{0.6}(r) D(y)
f(y)
\Bigl[ \frac{2\ell^2+2\ell-1}{(2\ell-1)(2\ell+3)} j_{\ell}(ky)
\nonumber \\ & &
- \frac{\ell(\ell-1)}{(2\ell-1)(2\ell+1)}j_{\ell-2}(ky)
\nonumber \\ & &
- \frac{(\ell+1)(\ell+2)}{(2\ell+1)(2\ell+3)}j_{\ell+2}(ky)
\Bigr]\,\rmd y.
\end{eqnarray}
Here $b$ is the standard Gaussian bias, while $\Delta b$ is a correction that applies to contributions from the curvaton field {\em only}:
\beq
\Delta b = 3 \frac{x_1}{x_2} (b-1) \delta_c \frac{\Omega_m}{k^2T(k)D(r)} 
\left(\frac{H_0}{c}\right)^2.
\label{eq:DB}
\eeq

\section{Constraints}
\label{sec:constraints}

To constrain $x_1$ and $x_2$ parameters using large scale structure we use the code developed and first implemented in Slosar {\slshape et~al.} \cite{Slosar08}.
We included the same data: the 5-year WMAP bispectrum $x_1=51\pm 30$ (1$\sigma$) \cite{Komatsu08}; and the SDSS data (spectroscopic and 
photometric luminous red galaxies, the photometric quasar sample, and the integrated Sachs-Wolfe effect via cross-correlation).
We further included ancillary data to constrain the background cosmological model and break degeneracies with the non-Gaussianity
parameters $(x_1,x_2)$: the CMB power spectrum \cite{p1,p2,p3,p4,p5} and supernova data \cite{p6}.

\begin{figure}
\includegraphics[width=3.3in]{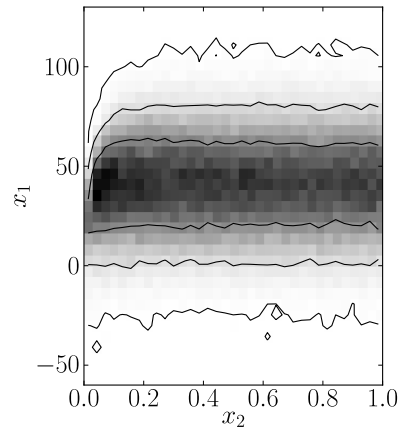}
\caption{\label{fig:2dplot}Constraints in the $(x_1,x_2)$ plane, including both the CMB bispectrum and the galaxy power spectrum.}
\end{figure}

The Markov chain results are displayed in Fig.~\ref{fig:2dplot} where the probability density distribution is plotted on the $(x_1,x_2)$ plane. Dark regions show regions with the highest likelihood and 
the contours outlines $68\%$, $95\%$ and $99.7\%$ confidence levels. As the role of the curvaton field in the primordial density perturbation decreases, i.e. $x_2 \rightarrow 0$ ($\xi\rightarrow\infty$), 
the upper limit on $x_1$ 
decreases.  This is because at small $x_2$, LSS becomes much more sensitive to $x_1$, as one may see from the $x_2$ in the denominator of Eq.~(\ref{eq:DB}); the $|W_\ell^1(k)|^2$ term in 
Eq.~(\ref{eq:cgv}) scales as $x_1^2/x_2$.  In particular, if a local-type CMB bispectrum is ever robustly detected ($x_1\neq 0$) then the non-detection of excess large-scale clustering in SDSS would 
immediately set a lower limit on $x_2$.


\section{Discussion}
\label{sec:disc}

This paper has extended the analysis of non-Gaussianity constraints into a two field inflationary models. In most previous studies of non-Gaussianity it was assumed that primordial density perturbations 
were generated either by inflaton field, in which case they are perfectly Gaussian, or only by the second field (for example curvaton) which contains quadratic part and generates non-Gaussian initial 
conditions.  It is important, however, to realize the possibility of an intermediate case where part of the curvature perturbation is derived from quantum fluctuations of the inflaton field, while an 
additional part is associated with a second field and converted to an adiabatic perturbation upon its decay.  This results in a peculiar type of non-Gaussian initial condition (which we may call 
``local-stochastic'' since the field $\phi_2$ entering in the nonlinear term is correlated with but not identical to the linear potential) that is both observable and distinguishable from the 
curvaton-only ``local-deterministic'' or $f_{\rm NL}$ form.  This type of non-Gaussianity has two parameters: a nonlinear 
coupling coefficient $\fnl$, and the ratio $\xi$ of inflaton to curvaton contributions to the primordial density perturbation spectrum. We connect these parameters with parameters characterizing 
inflationary fields in Eqs.~(\ref{eq:fnl}) and (\ref{eq:xi}).

Using the power spectrum and bispectrum constraints from SDSS and WMAP we are able to constrain these parameters.  Adding two sets of constraints together allows us to break the degeneracy in the 
$(\fnl, \xi)$ parameters 
that exists with the CMB bispectrum alone.  If non-Gaussianity in the CMB is ever detected, and the bispectrum has the local configuration dependence, this will enable us to measure the relative 
contributions of the inflaton and curvaton.

We have found that in contrast to the local-deterministic non-Gaussianity, whose main effect on large scale structure is a scale-dependent increase in 
the bias, the local-stochastic non-Gaussianity can 
introduce stochasticity between the matter and halo distributions.  It can also lead to relative stochasticity between haloes of different masses, 
since Eq.~(\ref{eq:chi-k}) depends on the Gaussian bias $b_{\rm 
g}$ of the haloes (e.g. $\chi\rightarrow 1$ if $b_{\rm g}\rightarrow 0$).  The potential use of these effects to directly test the hypothesis of multiple fields contributing to the primordial 
perturbations is left to future work.

\section*{Acknowledgments}

The authors are grateful to Shirley Ho for providing the large scale structure samples used in this study. 
D. T. and C. H. are supported by the U.S. Department of Energy (DE-FG03-92-ER40701) and the National Science Foundation (AST-0807337).
C. H. is supported by the Alfred P. Sloan Foundation.
This work was supported in part by the U.S. Department of Energy under
Contract No. DE-AC02-98CH10886.


\bibliographystyle{prsty}

\end{document}